\documentclass[runningheads]{llncs}
\usepackage[T1]{fontenc}
\usepackage{graphicx}
\usepackage{amsmath}
\usepackage{amssymb}
\usepackage{algorithm}
\usepackage{algorithmic}
\usepackage{bm}
\usepackage{color}
\usepackage{svg}

\usepackage{etoolbox}

\makeatletter
\patchcmd{\@makefntext}{\insert\footins\bgroup}{\insert\footins\bgroup\let\@makefnmark\relax}{}{}
\makeatother

\newcommand{\footfirstpage}[1]{
  \begingroup
  \renewcommand\thefootnote{} % 重定义脚注编号为空
  \footnotetext{#1} % 添加脚注文本
  \endgroup
}

\begin{document}
%
% \title{Versatile Experiment Designer Based on Human Concept Representations}
\title{CoCoG-2: Controllable generation of visual stimuli for understanding human concept representation}
\titlerunning{Controllable generation of visual stimuli in concept space}
% \author{Anonymous}
% %
% \authorrunning{Anonymous et al.}
\author{
Chen Wei$^{\dagger,1,2}$\and
Jiachen Zou$^{\dagger,1}$\and
Dietmar Heinke$^2$\and
Quanying Liu$^{*,1}$\\}
\institute{
$^1$Southern University of Science and Technology, Shenzhen, China \\
$^2$University of Birmingham, Birmingham, United Kingdom\\}

% \institute{Anonymous}
%
\maketitle              % typeset the header of the contribution
\footfirstpage{$\dagger$ Equal contribution.}
\footfirstpage{$^*$ Corresponding author.}

\begin{abstract}

Humans interpret complex visual stimuli using abstract concepts that facilitate decision-making tasks such as food selection and risk avoidance. Similarity judgment tasks are effective for exploring these concepts. However, methods for controllable image generation in concept space are underdeveloped.
In this study, we present a novel framework called CoCoG-2, which integrates generated visual stimuli into similarity judgment tasks. CoCoG-2 utilizes a training-free guidance algorithm to enhance generation flexibility. 
CoCoG-2 framework is versatile for creating experimental stimuli based on human concepts, supporting various strategies for guiding visual stimuli generation, and demonstrating how these stimuli can validate various experimental hypotheses. 
CoCoG-2 will advance our understanding of the causal relationship between concept representations and behaviors by generating visual stimuli.
The code is available at \url{https://github.com/ncclab-sustech/CoCoG-2}.

\keywords{Concept representation \and Conditional generative model \and Experimental design.}
\end{abstract}

\section{Introduction}

% Intro
% 1.1 什么是概念表征. (ref cocog intro para 1
% 1.2 通过相似度判断任务来探索概念表征，得到concept embedding. (ref cocog intro para 3
% 1.3 使用AI模型来探索概念表征. (ref cocog intro para 3
% 1.4 缺乏探索的方向：基于概念表征生成视觉刺激. (ref cocog intro para 1&3

% 人类在日常生活中会接收丰富的视觉刺激，并将其编码为功能、可食用性、生物性等抽象概念。人类基于视觉对象的概念表征执行择食、避险等高级认知决策行为。大量的人类研究表明，相似性判断任务是揭示人类概念表征的有效实验范式，而视觉对象概念表征之间的相似度也是各种高级认知任务的决策基础~\cite{roads2023modeling,hebart2020revealing}。概念空间中视觉对象之间的距离反映了人类头脑中视觉对象的距离，人类决策行为可以通过概念表征的不同维度来解释，而相似度判断则有望反映概念表征的所有维度。
% ----为了理解概念表示和行为之间的因果关系，需要关于概念表征的人类决策数据。因此有必要操纵概念，保留所有其他低级特征，以生成视觉对象，从而产生实验试次。这是人工智能鲜有探索的领域，即\textit{基于概念表示的可控视觉对象生成}。
% 以往研究显示，通过控制特定实验输入，可以直接观察影响决策的因素，从而更准确地建立因果关系，即反事实解释。同理，通过调整实验刺激中对象的概念表征，我们有望更好地理解概念表征与行为之间的因果联系。然而，由于概念通常隐藏于复杂的自然图像刺激之下，难以进行精确操纵，这一领域的研究仍然较少。

Humans receive abundant visual stimuli in their daily lives. By understanding visual input's abstract concepts such as function, edibility, and biological properties, humans can perform high-level cognitive decision-making tasks such as food selection and risk avoidance. 
A large number of human studies have shown that the similarity judgment task is an effective experimental paradigm for revealing human concept representation~\cite{murphy1985role,medin1993respects,roads2023modeling,hebart2020revealing}. In the similarity judgment task, participants are required to evaluate various visual stimuli and make judgments based on the similarity of these images.
Such similarity is measured in the concept space, rather than the original image space. So their distance in the concept space reflects the psychological space in the human mind~\cite{medin1993respects,roads2023modeling}. The distance in main dimensions of concept representation can predict and explain human decision-making behaviors; in turn, similarity judgments are expected to reflect key dimensions of concept representation.
Previous studies have demonstrated that by controlling specific experimental inputs, we can directly observe the core factors that influence decision-making, a process referred to as counterfactual reasoning~\cite{peterson2021using,fernandez2020explaining}. Similarly, by manipulating the concept representation within experimental stimuli, we can better uncover and understand the causal relationship between concept representations and behaviors. However, The concepts of complex natural visual stimuli often remain obscured and are quite challenging to manipulate. How to reveal the underlying concept space and generate the image by manipulating specific concepts remains largely unexplored.

\begin{figure*}[!t]
\centering
\includegraphics[width=12cm]{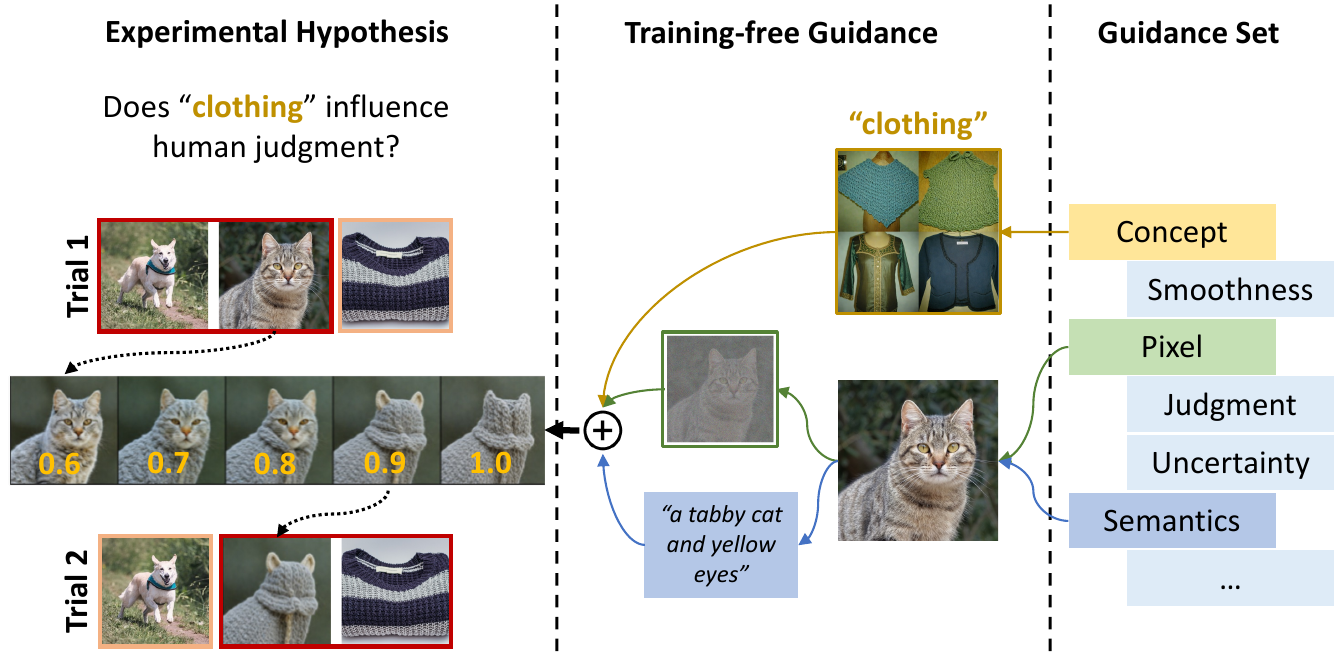}

\caption{Guided by the CoCoG-2, we can validate a specific hypothesis by generating visual stimuli in the controllable concept space using our framework. For instance, researchers propose an experimental hypothesis (\textit{Does "clothing" influence human judgment?}), and then construct a loss function using a guidance set (aimed at preserving pixel and semantic features while modifying the concept of "clothing"), and finally generate alternative visual stimuli through training-free guidance (visual stimuli where the concept of "clothing" exceeds 0.9 lead to changes in human judgment). These synthetic images can be used to test the hypothesis with behavioral experiments .}

\label{fig:framework}
\end{figure*}

%%%%%%%%%%%
% 2. 当前的方法有哪些。
% 2.1 简述可控生成模型（ref cocog intro para 2
% 2.2 基于人类反馈的可控生成模型（ref cocog intro para 2
% 2.3 ---可控生成模型辅助实验设计 
% 引入CoCoG

% 另一方面，可控生成模型的进步已经显着塑造了AI领域，尤其是通过使用gan和扩散技术的有条件生成模型〜\ cite {goodfellow2020 generative，tao20222df，song2020score，ho20202020denoising，ho2loisising，ho2n222222classifier}。 这些模型的应用涵盖了各种领域，例如文本到图像/视频综合，图像恢复和生物医学成像 生成条件来自各种来源，包括文本，草图和边缘，分割或深度〜\ cite {rombach2022high，Ramesh2022Hierarchical，meng2021sdedit，zhang2023Adding，YU2023Adding，Yu2023freedom，Bansal20223universal}。 尽管有这些进展，但当前的模型缺乏人类主观输入（例如感觉或反馈）的整合，通常会导致输出与人类偏好不符。
% 为了更好地满足人类的期望，最近的研究将人类的反馈纳入了这些模型，包括视觉偏好得分和人类中的比较中的决策〜\ cite {wu2023human，kirstain2023pick，von2023pick，von2023fabric，fan2023dpok，tang2023meroth}。 但是，这些举措尚未将认知科学的见解，尤其是人类决策过程中最具影响力的因素，即概念，纳入图像产生的控制变量。

Progress in controllable generation models has significantly shaped the AI field, especially through conditional generation models utilizing GANs and diffusion techniques~\cite{goodfellow2020generative,tao2022df,song2020score,ho2020denoising,ho2022classifier}. Applications of these models span various domains such as text-to-image/video synthesis, image restoration, and biomedical imaging~\cite{rombach2022high,ramesh2022hierarchical,kawar2022denoising,chung2022improving,meng2021sdedit,song2021solving,ozbey2023unsupervised}. The generation conditions derive from diverse sources including text, sketches, and maps of edges, segmentation, or depth~\cite{rombach2022high,ramesh2022hierarchical,meng2021sdedit,zhang2023adding,yu2023freedom,bansal2023universal}. Despite these advances, the current models lack integration of human subjective inputs such as feelings or feedback, often resulting in outputs that do not align with human preferences.
To better meet human expectations, recent studies have incorporated human feedback into these models, including visual preference scores and decisions from human-in-the-loop comparisons~\cite{wu2023human,kirstain2023pick,von2023fabric,fan2023dpok,tang2023zeroth}. 
Furthermore, it is important to integrate human cognitive mechanisms, such as decision-making based on concepts, into generative models. This integration can make the models more human-like and uncover human concept representations by manipulating concepts in generated images.

%%%%%%%%%%%
% 3.1 简述CoCoG，突出其贡献（使用概念表征生成可控的视觉刺激。特别的，CoCoG初步提出了通过生成视觉刺激高效探索概念表征，从而为人类行为提供counterfactual explanations）！
% 3.2 CoCoG的缺点。
% CoCoG只能输入完整的concept embedding作为引导的条件，这导致了引导生成缺乏灵活性。例如：（ref 如下

% 在先前研究中，Wei et al.提出了一个基于概念的可控发电（CoCoG）框架。 CoCoG 利用概念嵌入作为图像生成模型的条件，从而通过操纵概念表征来间接调控行为。 CoCoG 包含两部分：一个用于通过预测人类行为来学习概念嵌入的 \textit{概念编码器} 和一个 \textit{概念解码器}，它采用条件扩散模型通过两阶段生成策略将概念嵌入映射到视觉刺激。
% CoCoG 的概念编码器可以准确地预测人类视觉相似性决策行为，揭示了可靠的、可解释的人类的概念表征空间。CoCoG 的概念解码器可以通过控制概念嵌入来生成视觉刺激。 生成的视觉刺激与目标概念嵌入具有高度一致性，并可以通过操控概念维度来调控人类相似度决策行为。特别的，CoCoG初步提出了通过生成视觉刺激高效探索概念表征，从而为人类行为提供counterfactual explanations。
% 然而，CoCoG只能输入完整的concept embedding作为引导的条件，这导致了引导生成缺乏灵活性。一方面，CoCoG无法在保持图像其他特征不变的情况下对图片的concept进行编辑，且concepts之间可能产生冲突。另一方面，CoCoG无法直接根据预期的similarity judgment去生成图像，需要手动地挑选“key concept” (ref fig. ?) 去引导图片生成。这些都限制了该方法在进一步的基于概念表征的认知研究中的应用。

To generate images by controlling the concept space, a pioneer study, Wei et al.~\cite{wei2024cocog}, has proposed a concept-based controllable generation (CoCoG) framework. CoCoG utilizes concept embeddings as conditions for the image generation model, thereby indirectly manipulating behaviors through concept representations. CoCoG comprises two parts: a concept encoder that learns concept embeddings by predicting human behaviors, and a concept decoder that maps concept embeddings to visual stimuli using a conditional diffusion model with a two-stage generation strategy.
The concept encoder in CoCoG can accurately predict human visual similarity decision-making behaviors, revealing a reliable and interpretable space of human concept representations. The concept decoder can generate visual stimuli controlled by concept embeddings. The generated visual stimuli are highly consistent with the target concept embeddings and can regulate human similarity decision-making behaviors by manipulating concept dimensions. Notably, CoCoG preliminarily proposes efficiently exploring concept representations through generated visual stimuli, thus providing counterfactual explanations for human behavior.
However, CoCoG has its limitations. CoCoG can only input complete concept embeddings as guiding conditions, leading to a lack of flexibility in guided generation. Firstly, CoCoG cannot edit the concepts of an image while keeping other image features unchanged, and conflicts may arise between concepts. Secondly, CoCoG cannot directly generate images based on expected similarity judgments and requires manually selecting "key concept" to guide image generation. These limitations restrict the method's further application in cognitive research based on concept representations.

%%%%%%%%%%%
% 4. CoCoG-2的优势和贡献
% （如下
% 鉴于CoCoG框架的不足，我们提出了CoCoG-2框架，一个基于Concept based Controllable Generation的Versatile Experiment Designer。CoCoG-2 可以直接利用目标概念维度或是行为结果来引导图像扩散生成过程，使得高效设计和产生实验试次成为可能。CoCoG-2 对CoCoG 的两阶段生成架构做出了两大改进：在Prior Diffusion阶段，但不同于CoCoG使用条件扩散模型，CoCoG-2使用了Training-free Guidance方法，在目标概念维度或是行为结果的引导下，使用各种精心设计的guidance生成clip embedding；在图像生成阶段，CoCoG-2使用了img2img方法，从而进一步保持了图像的低维特征。
% 本文的贡献
% 1. 提出了通过training-free guidance设计基于人类概念表征设计实验并生成视觉刺激的通用流程。
% 2. 验证了多种潜在的guidance方法用于constrain视觉刺激，包括concept guidance, Smoothness guidance...
% 3. 实验结果表明通过组合不同的guidance产生的视觉刺激可以满足多种实验目的，丰富我们探索概念表征的工具。

In this study, we upgrade the CoCoG framework to CoCoG-2. CoCoG-2 allows us to design experimental stimuli based on concepts and effectively validate hypotheses about concept representation. Moreover, we employ a training-free guidance algorithm for the controllable generation of diffusion models with high flexibility. Our work has three main contributions.
\begin{itemize}
\item We present a general framework for designing experimental stimuli based on human concept representations and integrating experimental conditions through training-free guidance.
\item We have verified a variety of potential guidance strategies for guiding the generation of visual stimuli, controlling concepts, behaviors, and other image features.
\item Our experimental results demonstrate that visual stimuli generated by combining different guidance strategies can validate a variety of experimental hypotheses and enrich our tools for exploring concept representation.
\end{itemize}

\section{Preliminaries}

\subsection{Training-free Guidance Diffusion Models}

\subsubsection{Diffusion Models.}

Diffusion models~\cite{song2020score,karras2022elucidating} operate through two distinct phases: the forward and the reverse processes. During the forward phase, which spans from time $0$ to $T$, the model progressively converts an image into Gaussian noise. Conversely, during the reverse phase, it reconstructs the image starting from noise, reversing the timeline from $T$ back to $0$. Denote $x_t$ as the state of the data at time $t$. In the forward phase, noise is methodically added according to a specific noise schedule:

\begin{equation}
 x_t = a_t x_0 + b_t \epsilon_t, 
\end{equation}

where, $a_t = \sqrt{\alpha_t}, b_t = \sqrt{1 - \alpha_t}$, $ \alpha_t $ varies monotonically from 0 to 1 as $ t $ increases, and $ \epsilon_t $ is sampled from a standard normal distribution $ \mathcal{N}(0, I) $. The neural network in diffusion models is trained to approximate the noise at each timestep:

\begin{equation}
\min_{\theta} \mathbb{E}_{x_t, \epsilon_t} \left[ \| \epsilon_{\theta}(x_t, t) - \epsilon_t \|_2^2 \right] = \min_{\theta} \mathbb{E}_{x_t, \epsilon_t} \left[ \| \epsilon_{\theta}(x_t, t) + b_t \nabla_x \log p_t(x_t) \|_2^2 \right],
\end{equation}

where $ p_t(x_t) $ represents the probability distribution of $ x_t $. The dynamics of the reverse phase are governed by the following ordinary differential equation (ODE):

\begin{equation}
 \frac{dx_t}{dt} = f(t)x_t - \frac{g^2(t)}{2} \nabla_x \log p_t(x_t)
\end{equation}

with $ f(t) = -\frac{d \log a_t}{dt} $ and $ g^2(t) = \frac{d b^2_t}{dt} - 2\frac{d \log \sqrt{\alpha_t}}{dt} b^2_t $. Through these equations, the reverse process manages to transform Gaussian noise back into coherent images.

In practical applications, the update rules for both the forward and reverse processes depend on the choice of the sampling algorithm. In this work, we employ the DDPM algorithm~\cite{ho2020denoising} and we represent the forward and reverse steps using the following notations respectively:

\[
x_t = DDPM^{+}(x_{t-1}) \quad \text{and} \quad x_{t-1} = DDPM^{-}(x_t).
\]

\subsubsection{Training-Free Guidance.}

In diffusion models, conditional generation introduces a specific condition \( y \) into the generative process~\cite{song2020score,ho2022classifier}. By applying Bayes' rule \( p(x|y) = \frac{p(y|x) p(x)}{p(y)} \), it incorporates the condition through an additional likelihood term \( p(x_t|y) \):

\begin{equation}
\nabla_{x_t} \log p_t(x_t | y) = \nabla_{x_t} \log p_t(x_t) + \nabla_{x_t} \log p_t(y | x_t).
\label{eq:score function}
\end{equation}

Recent approaches, known as training-free methods~\cite{chung2022improving,bansal2023universal,yu2023freedom,shen2024understanding,yang2024guidance}, utilize pre-trained diffusion models for conditional generation without the need for retraining or finetuning. These methods employ a pre-trained diffusion model and a differentiable loss function \( \ell(f_{\phi}(x_0), y) \), defined using a neural network \( f_{\phi} \). Tweedie’s formula is used to compute \( \nabla_{x_t} \log p_t(y | x_t) \) by estimating \( \hat{x}_0 \) from \( x_t \):

\begin{equation}
\hat{x}_0(x_t) \approx \mathbb{E} [x_0 | x_t] = \frac{x_t - b_t \epsilon_t(x_t, t)}{a_t}.
\end{equation}

\begin{equation}
\nabla_{x_t} \log p_t(y | x_t) = \nabla_{x_t} \ell(f_\phi(\hat{x}_0(x_t)), y),
\label{eq:loss function}
\end{equation}

Then, we can add an additional correction step in the reverse sampling process:

\[
x_{t-1} = DDPM^{-}(x_t) - \eta \nabla_{x_t} \ell(f_\phi(\hat{x}_0(x_t)), y)
\]

This approximation allows the use of an existing network, which is initially trained for processing clean data ($x_0$). The gradient of the final term in the loss function is determined through backpropagation across the guidance network and the diffusion framework. In our example, where the condition is concept dimension, $f_{\phi}$ is modeled using a concept encoder, and $\ell$ is the loss designed for the experiment.

\subsection{Concept based Controllable Generation}
\label{section:cocog}

The CoCoG framework, introduced by Wei et al.~\cite{wei2024cocog}, utilizes concept embeddings as conditional inputs for image generation, bridging cognitive science with artificial intelligence. CoCoG aims to enhance the production of natural visual stimuli in a controllable manner, closely aligning with human cognitive processes. It comprises two key components: a concept encoder for learning concept embeddings through human behavior prediction, and a concept decoder for transforming these embeddings into visual stimuli via a conditional diffusion model.

\subsubsection{Concept Encoder}
The concept encoder extracts concept embeddings from visual objects. Each image $x$ from a dataset is processed by the CLIP image encoder $f$ to produce a CLIP embedding $h$. This embedding is then transformed by a learnable projector $g$ into a concept embedding $c$:

\begin{equation}
\begin{aligned}
&\text{CLIP embedding:} & h = f(x),\\
&\text{concept embedding:} & c = g(h),
\end{aligned}
\end{equation}

Each dimension of $c$ corresponds to an interpretable concept, with its activation value reflecting the concept's prominence in the visual object. Training is conducted using the {\it triplet odd-one-out} task from the THINGS dataset~\cite{hebart2023things}, employing dot product similarity and cross-entropy for decision prediction:

\begin{equation}
p(y) = Softmax(S_{jk}, S_{ik}, S_{ij}),
\label{eq:similarity}
\end{equation}

Here $S_{ij} = <c_i, c_j>$, and predicted behavior $y$ are matched against human judgments to train the projector $g$, with $L_1$ regularization ensuring sparsity in the concept embedding $c$.

\subsubsection{Concept Decoder}
Following encoder training, triplets $(x_i, h_i, c_i)$ are used by the concept decoder to control the generation of visual objects based on human concept representation. The process is formalized as:

\begin{equation}
p(x, h, c, y) = p(y)p(c|y)p(h|c)p(x|h).
\label{eq:2-stage}
\end{equation}

The decoder operates in two phases. In stage I (prior diffusion), conditioned on $c$, a diffusion model learns the distribution of CLIP embeddings $p(h|c)$ using a lightweight U-Net $\epsilon_{prior}(h_t, t, c)$. The model trains on ImageNet pairs $(h_i, c_i)$ using classifier-free guidance.
In Stage II (CLIP-guided generation), with the CLIP embedding $h$ from Stage I, a generator models $p(x|h)$ using pre-trained SDXL and IP-Adapter models~\cite{podell2023sdxl,sauer2023adversarial,ye2023ip}, which facilitate the integration of $h$, to guide the U-Net's denoising process.

\subsubsection{CLIP Embedding as an Intermediate Variable} 
They use CLIP embeddings as an intermediate variable for two reasons:
1) CLIP embeddings are low-level and retain key image information, effectively predicting human similarity judgments with linear probing~\cite{muttenthaler2022human}.
2) They enable the use of pre-trained generative models, adopting a two-stage generation strategy that requires only training a Prior Diffusion model, significantly reducing computational costs.

Despite its efficacy in generating controllable visual stimuli, CoCoG's reliance on full concept embeddings for input restricts flexibility. It cannot modify specific concepts without altering other image attributes, leading to potential conflicts. Furthermore, it cannot generate images directly from predicted similarity judgments (i.e., $p(x|y)$), requiring manual selection of key concepts for image generation (i.e., $p(x|c=c(y))$), which limits its utility in detailed cognitive research. To address these limitations, further work should explore methods that allow for more flexible manipulation of specific concepts while maintaining control over other image attributes. For instance, generating multiple trials with consistent visual stimuli or varying only specific concepts without altering other features could be beneficial for more complex experimental designs. This improvement, which will be discussed below, would enhance CoCoG's applicability in diverse research contexts.

\section{Method}

% 相对于CoCoG，CoCoG-2改进了concept decoder的controllable generation策略。我们将需要建模的分布从\ref{eq:2-stage}简化为$p(x, h, e) = p(e)p(h|e)p(x|h)$. 其中$e$表示我们希望visual stimuli需要达到的条件，包括concept $c$, similarity judgment $y$ and others.

Compared to CoCoG, CoCoG-2 introduces improvements in the controllable generation strategy of the concept decoder. We simplify the distribution model from Equation \ref{eq:2-stage} to $p(x, h, e) = p(e)p(h|e)p(x|h)$. Here, $e$ represents the conditions that the visual stimuli need to be controlled, including concepts $c$, similarity judgments $y$, and others.

% 1. 使用training-free guidance提示
% 在CoCoG-2的concept decoder中，我们依然使用两阶段策略，即先建模p(h|e)，再建模p(x|h)。
% 在这篇工作中，我们尝试通过在建模时p(h|e)引入training-free guidance提升CoCoG的灵活性和应用范围。
% 如上一章节所诉，直接学习分布p(h|c)的策略缺乏灵活性。
% 我们选择只学习先验分布p(h)，同时使用公式\ref{eq:score function}将p(h|e)分解为先验p(h)和似然p(e|h)。如公式\ref{eq:loss function},对于似然项p(e|h)，我们有$\nabla_{h_t} \log p_t(e | h_t) = \nabla_{h_t} \ell(f_\phi(\hat{h}_0(h_t)), e)$。
% 我们只需要根据实验目的，来选择合适的条件$e$和关于$h$的可导的损失函数$\ell(f_\phi(\cdot), \cdot)$，就可以有效的constrain visual stimuli （如图二）.

\subsection{Training-free guidance for controlling visual stimuli}

In the concept decoder of CoCoG-2, we continue to employ a two-stage strategy, first modeling $p(h|e)$ and then $p(x|h)$. In this work, we attempt to enhance the flexibility and applicability of CoCoG by introducing training-free guidance in the modeling of $p(h|e)$. As discussed in the Section~\ref{section:cocog}, learning distribution $p(h|c)$ directly lacks flexibility.
For training-free guidance, we only learn the prior distribution $p(h)$ and decompose $p(h|e)$ into the prior $p(h)$ and the likelihood $p(e|h)$ using Equation \ref{eq:score function}. As shown in Equation~\ref{eq:loss function}, for the likelihood term $p(e|h)$, we have $\nabla_{h_t} \log p_t(e | h_t) = \nabla_{h_t} \ell(f_\phi(\hat{h}_0(h_t)), e)$.
By simply choosing the appropriate condition $e$ and a differentiable loss function $\ell(f_\phi(\cdot), \cdot)$ based on the experimental hypothesis, we can effectively guide the generation of visual stimuli as demonstrated in Figure~\ref{fig:framework}.

% 2. Select guidance to construct the loss function
% 在实验设计中，我们往往希望visual stimuli同时满足多个条件。我们可以通过组合我们的不同需求来构造整体的损失函数$\ell = \sum \lambda_k \ell_k$，其中\lambda_k控制每种损失的权重。具体地，我们在实验中将讨论如下的种类多guidance以及构造他们损失函数的方法。

% Concept guidance: 如果我们希望生成的图片的concept达到预期的值，我们可以定义损失函数为：
% $\ell = \sum \| g(h)_i - c_i \|$
% 其中 i 为我们希望控制的concepts的编号，c_i是其目标值。我们可以一次控制个别的concepts，也可以控制所用的concepts（就想CoCoG）

% Smoothness guidance: 如果我们希望生成一批图片保持语义上的平滑，我们可以定义损失函数为：
% $\ell = \sum_{i,j \in S} \| h_i - h_j \|$
% 这里的 S = {i, j} 包含生成这一批图片中我们希望保持平滑的图片的编号。例如我可以生成一行两两相似的图片（如图四）。

% Semantics guidance: 使生成的图片和给定的图片语义上相似，我们可以定义损失函数为：
% $\ell = \| h - \bar{h} \|$
% 其中 \bar{h} 是给定的图片CLIP embedding。

% Judgment guidance: 如果希望人类在similarity judgment中的预期的行为概率分布服从目标分布，我们可以定义损失函数为：
% $\ell = -\sum p(y=i) log(\bar{p(y=i)})$
% 其中 \bar{p(y)} 是目标分布，该损失函数计算了预期分布和目标分布之间的交叉熵损失。由\ref{eq:similarity}可知E_{p(y)}[y]=Softmax \dot S \dot g (h),所以\ell依然是关于h的可导函数。

% Uncertainty guidance: 当我们需要控制similarity judgment的uncertainty时，我们可以定义损失函数为：
% $ell = H(p(y)) = -\sum p(y=i) log(p(y=i))$

% Outline guidance: 由于CLIP embedding $h$主要是控制生成图像的high-level特征或者说语义特征，我们也尝试引入img2img [citation] 来控制图像的low-level特征。

% concept; smooth; semantic; judgment; entropy; outline

\subsection{Guidance set and corresponding loss functions}

To construct the loss function, we often wish for the visual stimuli to satisfy multiple conditions via guiding the generative process. We can combine our various requirements to construct an overall loss function $\ell = \sum \lambda_k \ell_k$, where $\lambda_k$ controls the weight of each type of loss. Specifically, in our experiments, we will define the guidance set and their corresponding loss functions as follows:

\subsubsection{Concept guidance:} If we aim for the generated images to achieve predetermined concept values, we can define the loss function as:
\begin{equation}
\ell = \sum \| g(h)_i - c_i \|
\end{equation}
where \(i\) indexes the concepts we wish to control, and \(c_i\) are their target values. We can control individual concepts or all concepts simultaneously (Like CoCoG).

\subsubsection{Smoothness guidance:} If we desire a batch of images to maintain semantic smoothness, we can define the loss function as:
\begin{equation}
\ell = \sum_{i,j \in S} \| h_i - h_j \|
\end{equation}
Here, \(S = \{i, j\}\) includes the indices of images in the batch that we want to remain smooth. For example, we can generate a batch of pairwise similar images (as shown in Figure 4).

\subsubsection{Semantics guidance:} To make the generated images similar to a given image semantically, we can define the loss function as:
\begin{equation}
\ell = \| h - \bar{h} \|
\end{equation}
where \(\bar{h}\) is the CLIP embedding of the given image.

\subsubsection{Judgment guidance:} If we want human similarity judgments to follow a target judgment distribution, we can define the loss function as:
\begin{equation}
\ell = -\sum p(y=i) \log(\bar{p}(y=i))
\end{equation}
where \(\bar{p}(y)\) is the target distribution. This loss function calculates the cross-entropy loss between the expected and target distributions. Given by Equation \ref{eq:similarity}, \(E_{p(y)}[y] = Softmax \cdot S \cdot g(h)\), so \(\ell\) is still a derivable function with respect to \(h\).

\subsubsection{Uncertainty guidance:} When we need to control the uncertainty in similarity judgments, we can define the loss function using the entropy of the expected judgment distribution:
\begin{equation}
\ell = H(p(y)) = -\sum p(y=i) \log(p(y=i))
\end{equation}

\subsubsection{Pixel guidance:} Since CLIP embedding \(h\) primarily controls the high-level features or semantic features of the generated images, we also try to introduce img2img~\cite{meng2021sdedit} to control the low-level features of the images.

% 1.2 Improving training-free guidance
% 之前的研究显示，training-free guidance的生成过程不稳定，甚至容易出现guidance failure的情况。为了确保guidance的稳定和有效，我们分析了该领域近期的主要进展，employ two techniques to enhance training-free guidance。

\subsection{Improving Training-Free Guidance}

Previous research has indicated that the generation process of training-free guidance is unstable and prone to guidance failures. To ensure the stability and effectiveness of guidance, we analyzed recent major advancements in this area and employed two techniques to enhance training-free guidance.

\subsubsection{Adaptive gradient scheduling.}

Inspired by \cite{shen2024understanding,yang2024guidance}, the training guidance process can be viewed as optimizing the loss function of the guidance network. It stands to reason that adopting a more sophisticated optimizer could accelerate the convergence of the guidance. Following \cite{yang2024guidance}, we use a closed-form solution based on the spherical Gaussian constraint:

% \begin{equation}
% x_{t-1}^* = x_{t-1} - \eta \sqrt{n} \sigma_t \frac{\nabla_{x_t} \ell (f_{\phi}(\hat{x}_0), y)}{\| \nabla_{x_t} \ell (f_{\phi}(\hat{x}_0), y) \|_2},
% \end{equation}

\begin{equation}
h_{t-1}^* = h_t - \eta \sqrt{n} \sigma_t \frac{\nabla_{h_t} \ell (f_{\phi}(\hat{h}_0(h_t)), e)}{\| \nabla_{h_t} \ell (f_{\phi}(\hat{h}_0(h_t)), e) \|_2},
\end{equation}

% where $\sigma_t = \sqrt{(1 - \alpha_{t-1} / (1 - \alpha_t))} \sqrt{1 - \alpha_t / \alpha_{t-1}}$, $n$ is the dimension of the $x$.

where $\sigma_t = \sqrt{(1 - \alpha_{t-1} / (1 - \alpha_t))} \sqrt{1 - \alpha_t / \alpha_{t-1}}$, $n$ is the dimension of the $h$, and $\eta$ is the guidance scale.

\begin{algorithm}
% \centering
\caption{Improved training-free guidance for prior diffusion}
\begin{algorithmic}[1]
\FOR{$t = T, \ldots, 0$}

    % \FOR{$i = 1, \ldots, s$}

    %     \STATE $x_{t-1}^i = DDPM^{-}(x_t^i)$
    %     \STATE $\hat{x}_0^i = \frac{x_t^i - \sigma_t \epsilon_{\theta}(x_t^i, t)}{\sqrt{\alpha_t}}$
    %     \STATE $g_t = \nabla_{x_t} \ell (f_{\phi}(\hat{x}_0^i), y)$
    %     \STATE $x_{t-1}^i = x_{t-1} - \eta \sqrt{n} \sigma_t \frac{g_t}{\| g_t \|_2}$
    
    %     % \STATE $x_{t-1}^i =$ DDIM with Guidance($x_{t-1}^{i-1}$)
    %     \IF{$i < s$}
    %         \STATE $x_t^i = DDPM^{+}(x_{t-1}^i)$
    %         % \STATE $\beta_t = \alpha_t / \alpha_{t-1}, n \sim \mathcal{N}(0, I)$
    %         % \STATE $x_t^i = \sqrt{\beta_t} x_{t-1}^i + \sqrt{1 - \beta_t} n$
    %     \ENDIF
    %     \STATE $x_{t-1}^0 \gets x_{t-1}^s$
    % \ENDFOR
    
    \FOR{$i = 1, \ldots, s$}
    
        \STATE $h_{t-1}^i = DDPM^{-}(h_t^i)$
        \STATE $\hat{h}_0^i = \frac{h_t^i - \sigma_t \epsilon_{\theta}(h_t^i, t)}{\sqrt{\alpha_t}}$
        \STATE $g_t = \nabla_{h_t} \ell (f_{\phi}(\hat{h}_0^i(h_t^i)), e)$
        \STATE $h_{t-1}^i = h_{t-1} - \eta \sqrt{n} \sigma_t \frac{g_t}{\| g_t \|_2}$
    
        % \STATE $h_{t-1}^i =$ DDIM with Guidance($h_{t-1}^{i-1}$)
        \color{black} \IF{$i < s$}
            \STATE $h_t^i = DDPM^{+}(h_{t-1}^i)$
            % \STATE $\beta_t = \alpha_t / \alpha_{t-1}, n \sim \mathcal{N}(0, I)$
            % \STATE $h_t^i = \sqrt{\beta_t} h_{t-1}^i + \sqrt{1 - \beta_t} n$
        \ENDIF
        \STATE $h_{t-1}^0 \gets h_{t-1}^s$
    \ENDFOR
\ENDFOR

\end{algorithmic}
\end{algorithm}

\subsubsection{Resampling trick.}

Following \cite{yu2023freedom,shen2024understanding}, we use resampling, or ``time travel'', to address sampling errors. This involves reintroducing random noise to reset the sampling state, reducing distributional divergence and aligning samples with the target distribution.

\begin{figure*}[!h]
\centering
\includegraphics[width=12cm]{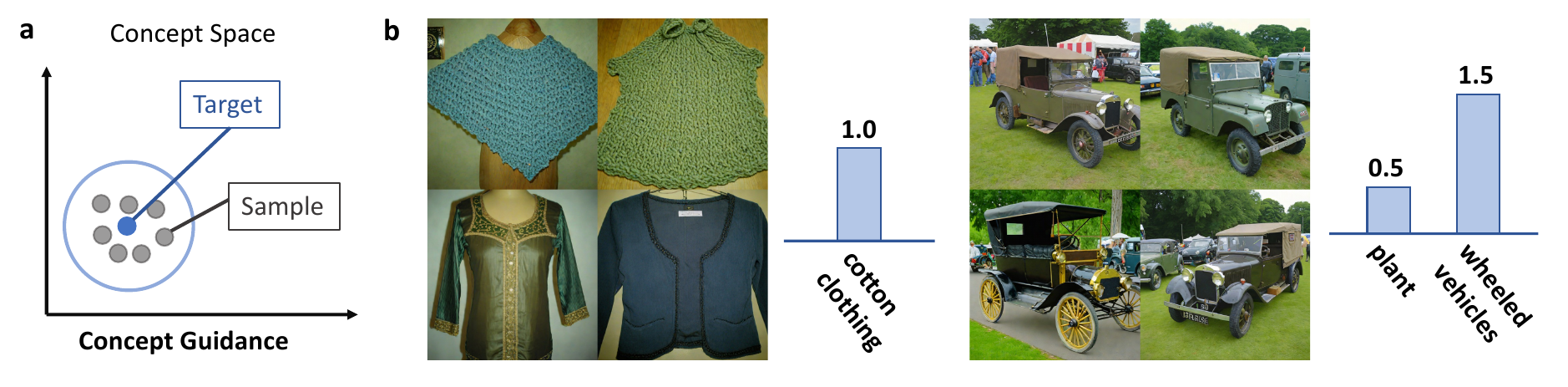}
\caption{Generate diverse visual stimuli for given target concepts. (a) Concept guidance used in this experiment. (b) Visual stimuli generated under the guidance of one or more target concepts.}
\label{fig:exp1}
\end{figure*}

\section{CoCoG-2 for versatile experiment design}

% run on an Nvidia 3090

% given ci -> mutiple x
\subsection{Diverse generation based on concept}

In the first experiment, we tested the effectiveness of using CoCoG-2 to generate visual stimuli based on given target concepts. The objective of the experiment was to generate diverse samples specific activation values of single or multiple concept dimensions. Therefore we only used \textit{Concept guidance} to guide the Prior Diffusion. As shown in Figure~\ref{fig:exp1}b, whether guided by a single concept or multiple concepts, CoCoG-2 was able to generate images that were well-aligned with the target concepts, and these images exhibited good diversity in features unrelated to given concept.

\begin{figure*}[!h]
\centering
\includegraphics[width=12cm]{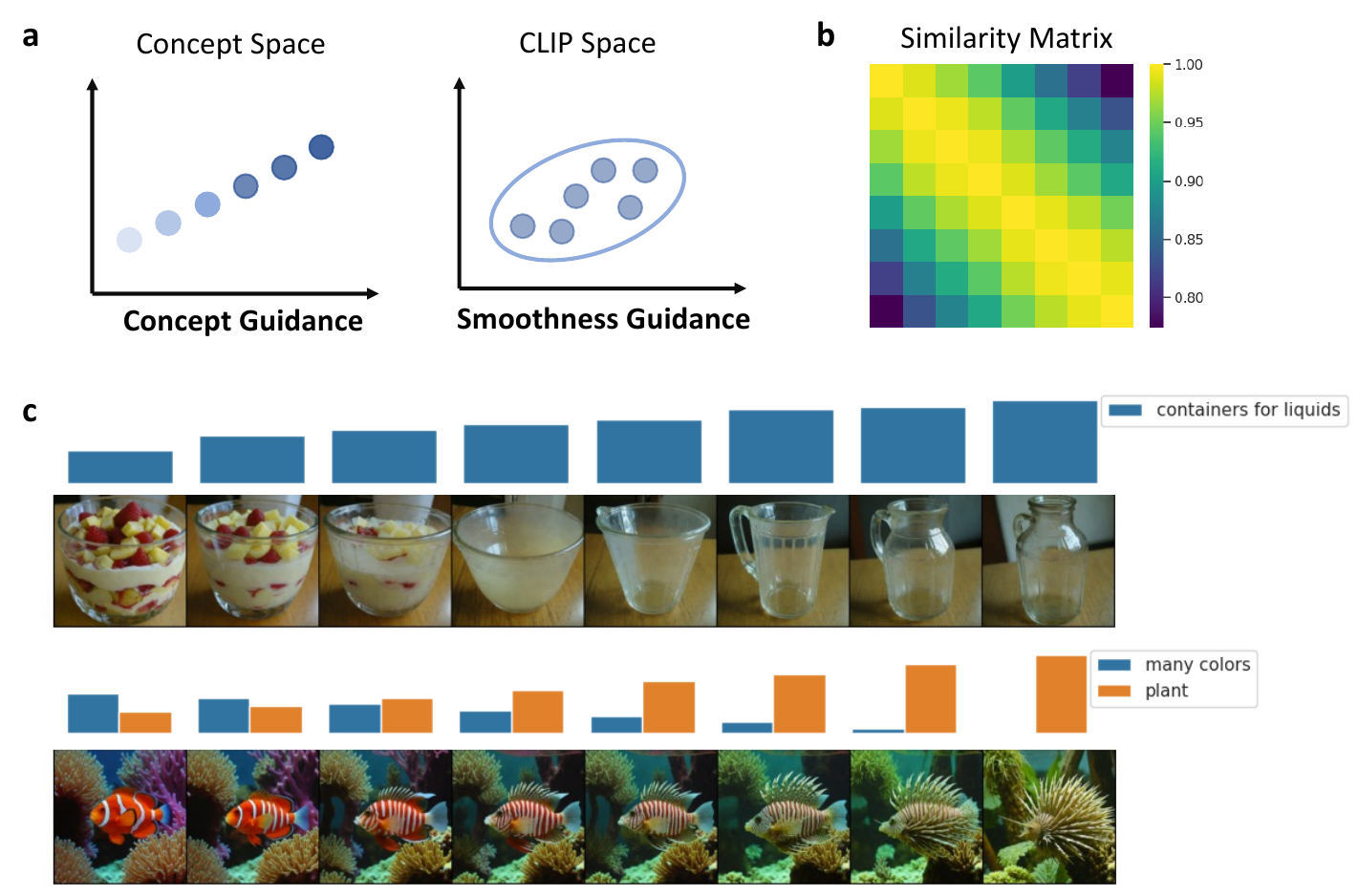}
\caption{The smooth change of visual stimuli generated based on concept. (a) Concept guidance and Smoothness guidance used in this experiment. (b) The average CLIP similarity matrix of 100 groups of stimuli. (c) Two trials of images generated under the guidance of one or more target Concepts.}
\label{fig:exp2}
\end{figure*}

% control ci -> auto prevent conflict
\subsection{Smooth changes based on concepts}

After confirming that target concepts can effectively guide the generation of visual stimuli, we modified the activation values of target concepts in the second experiment to generate a group of stimuli while maintaining smooth semantic features. We used \textit{Concept guidance} to allow gradual changes in concept activation values and employed \textit{Smoothness guidance} to maintain the similarity of semantic features between neighboring stimuli. We randomly generated 50 groups of visual stimuli and calculated the similarity matrix of CLIP embeddings for each group's stimuli, which was then averaged to obtain a mean similarity matrix. The similarity matrix shown in Figure~\ref{fig:exp2}b indicates that the CLIP similarity of images within a trial ranged between 0.75 and 1.0, with similarity increasing as the distance of image position decreased, consistent with the expected effect of Smoothness guidance. Figure~\ref{fig:exp2}c displays visual stimuligenerated under the guidance of one or more varying target concepts, showing clear and stable changes at the concept level, while other features were well maintained.

\begin{figure*}[!h]
\centering
\includegraphics[width=12cm]{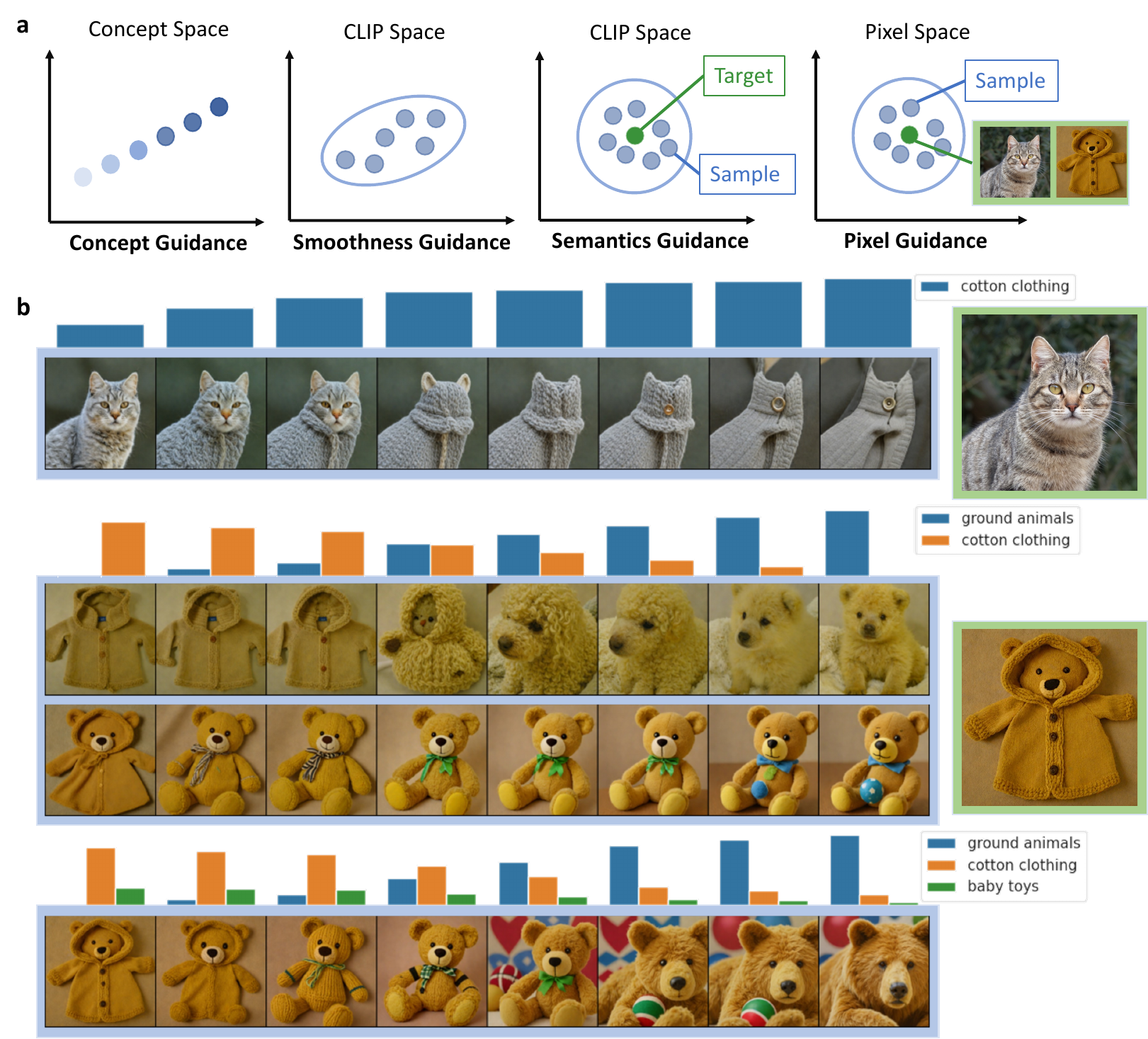}
\caption{Visual stimuli generated by image editing in concept. (a) \textit{Concept, Smoothness, Semantics}, and \textit{Pixel guidance} used in this experiment. (b) A trial of images generated under the guidance of a single concept (row 1); Visual stimuli generated by CoCoG-2 and CoCoG under the guidance of multiple concepts (row 2-4). When guided only by "ground animals" and "cotton clothing," the images generated by CoCoG-2 align well with the target concepts. However, the images generated by CoCoG cannot produce real animals due to conflicts between target concepts and the concept "baby toys" in the original image, necessitating manual adjustment of the "baby toys" value.}
\label{fig:exp3}
\end{figure*}

% given x, control ci -> img2img
\subsection{Image editing in concept}
% fig x_original (ci=ci), ci (0, 0.1, ...)

Next, we tested whether it was possible to start with an "original image" and edit target concepts to generate a group of stimuli that are similar to the original image and vary the concepts according to given values. For concept editing, we used \textit{Concept guidance}; to maintain other features of images within a trial, we applied \textit{Smoothness guidance}. Additionally, we utilized \textit{Semantics guidance} in CLIP Space and \textit{Pixel guidance} in Pixel Space to maximize the similarity between the generated images and the original image.

As shown in Figure~\ref{fig:exp3}b, Visual stimuli generated by CoCoG-2 not only resembled the original image but also achieved clear and stable concept edits, with well-preserved low-level features. This experiment also highlighted a significant advantage of CoCoG-2 over CoCoG: it can automatically avoid conflicts between target concepts and the concepts in the original image, without the need for manually modifying conflicting concepts. As shown in rows 2-3 of Figure~\ref{fig:exp3}b, both guided by "ground animals" and "cotton clothing", CoCoG-2 was able to generate real animals, whereas CoCoG could only generate fabric toy bears. This is because the concept "baby toys" in the original image conflicted with target concepts. Even if the activation value of "cotton clothing" was set to zero, the generated images still contained this concept. Therefore, as shown in row 4 of Figure~\ref{fig:exp3}b, it was necessary to manually modify the activation value of "baby toys" to generate real animals. 

CoCoG's manual modification process increased the complexity of visual stimuli generation, as conflicting concepts had to be identified and modified through multiple rounds of generation, and it is possible that contradictions could not be resolved by altering just a few key concepts. However, CoCoG-2’s Training-Free Guidance method could automatically adjust conflicting concepts, greatly speeding up the generation process.

\begin{figure*}[!h]
\centering
\includegraphics[width=12cm]{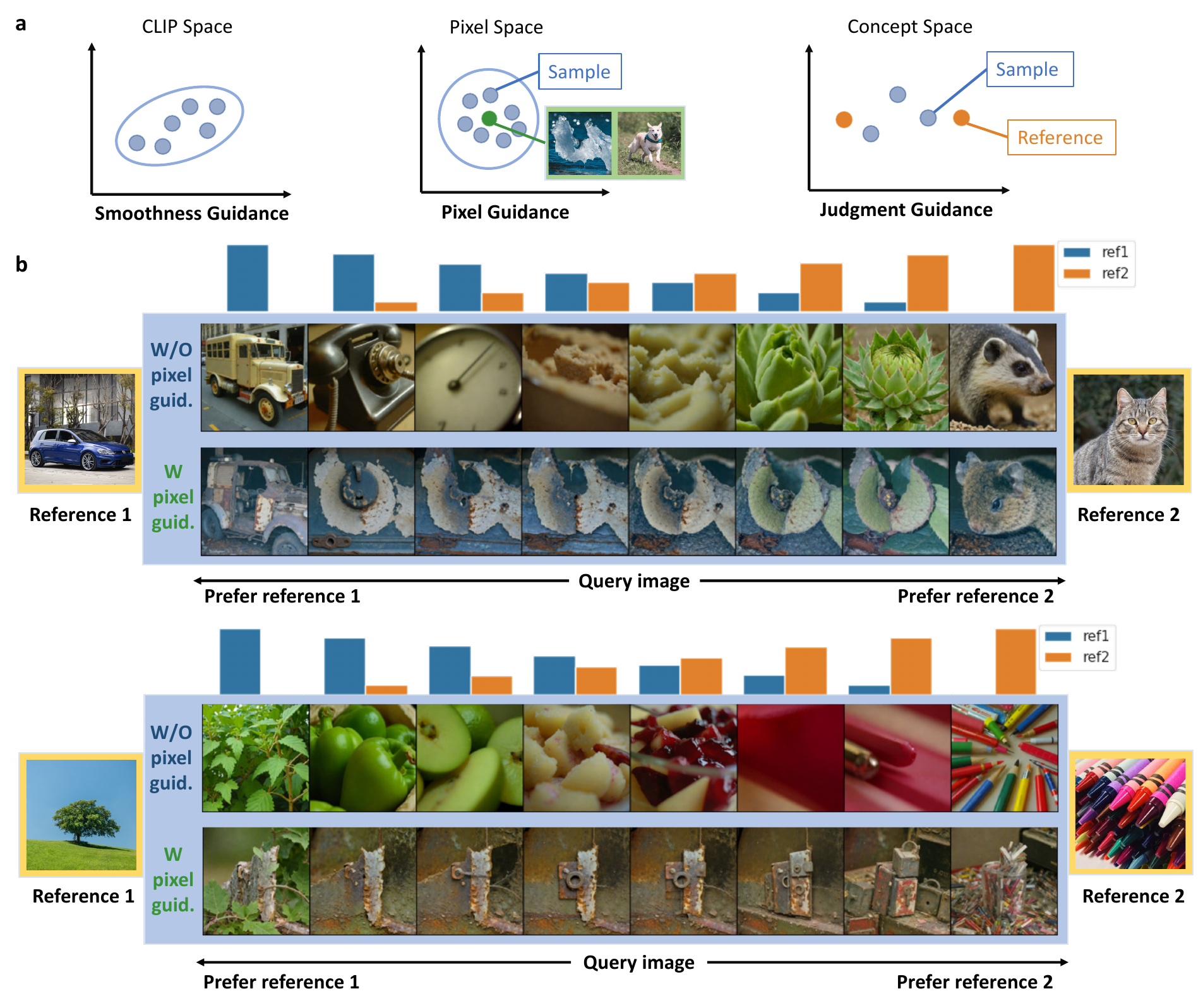}
\caption{Behavioral manipulation of similarity judgments with/without \textit{Pixel guidance}. (a) \textit{Smoothness, Pixel}, and \textit{Judgment guidance} used in this experiment. (b) Images generated under the guidance of probability interpolation of two pairs of references. The top row of images for each set is generated without using the \textit{Pixel guidance}, while the bottom row of images is generated with the \textit{Pixel guidance}.}
\label{fig:exp4}
\end{figure*}

% control y -> multiple paths
\subsection{Behavioral manipulation of similarity judgment}

After verifying CoCoG-2's capability of concept editing based on the original image, we further tested whether CoCoG-2 could directly generate visual stimuli guided by experimental results. We used the {\it Two alternative forced choice} experiment as an example, where participants need to choose between reference 1 \& 2, judging which is more similar to a query. In this scenario, the experimental results can be represented as the probability of choosing reference 1 or 2 as the more similar image. We used the probabilities of each reference image being chosen as the guidance, termed the {\it Judgment guidance}. Additionally, to study the causal relationship between visual cognitive behavior and concept representation, we needed to ensure that changes of behavioral results were solely caused by changes in concepts, which means maintaining other features of the images. Therefore, we used the {\it Smoothness guidance} to maintain other features within a group, and guided the query image's low-level features using an image unrelated to the reference images, termed the {\it Pixel guidance}.

In Figure~\ref{fig:exp4}b, we show the generated query images for two references. The upper group of each set is without the {\it Pixel guidance}, and significant variations in low-level features can be observed among the generated images. The lower group is with the Pixel guidance, where the images in Figure~\ref{fig:exp4}a labeled "ice" and "dog" were used to guide the low-level features respectively. Both groups of image align with the given results of similarity judgment. However, the lower group of images also maintains consistency with the guiding images in shape, color, and other low-level features. This greatly reduces the impact of low-level features on behavioral results, more reliably interpreting the causal relationship between behavior and concepts.

% \section{CoCoG-2 for cognitive research in human and AI}\label{sec:cognitive research}

\begin{figure*}[!h]
\centering
\includegraphics[width=12cm]{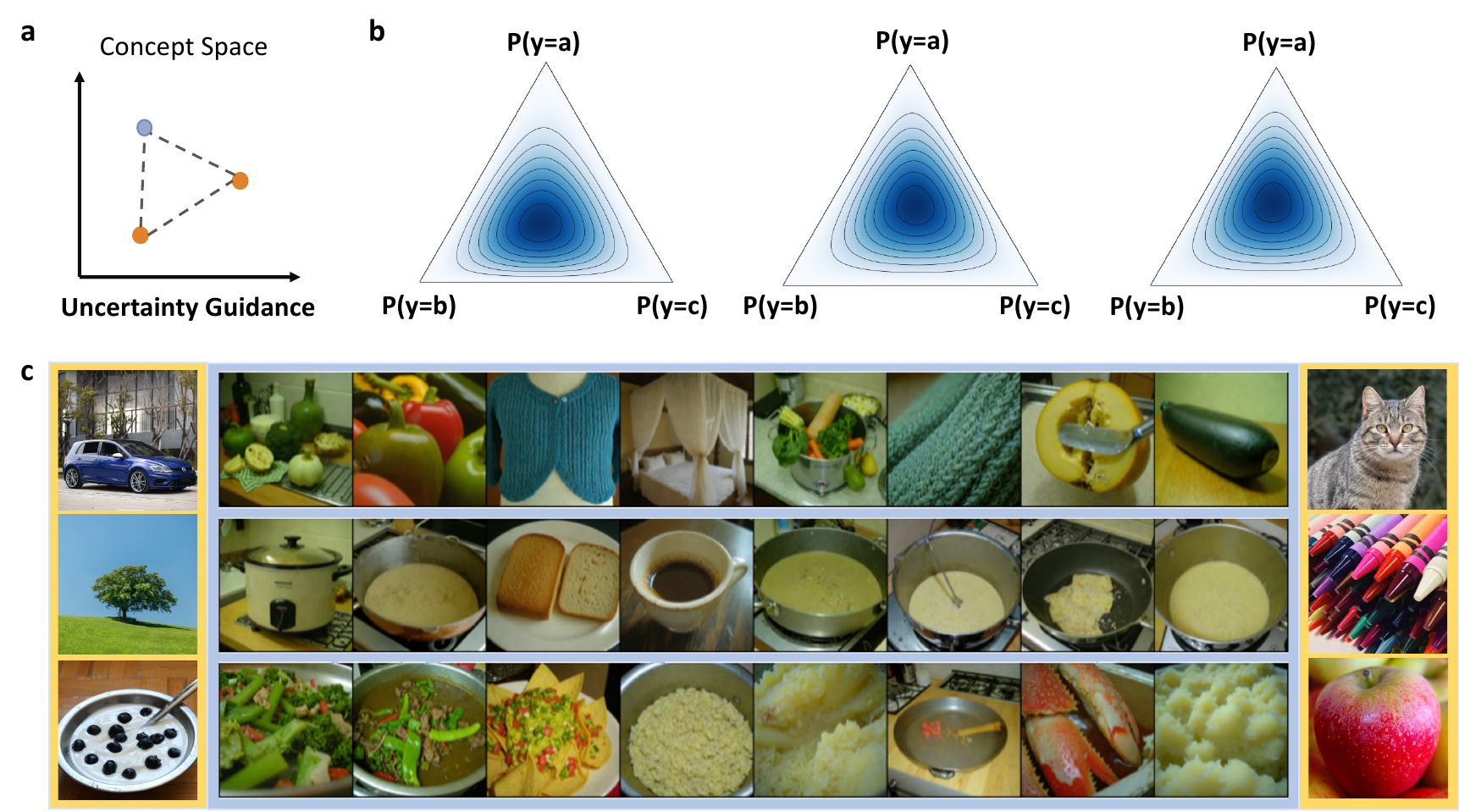}
\caption{Generated trials of similarity judgment leading to rich individual preference information. (a) {\it Uncertainty guidance} used in this experiment. (b) Probability distributions of three sets of generated images in c, with each side of the triangle representing the probability of choosing a particular image as the odd one out. (c) Images generated under the guidance of Uncertainty guidance for three pairs of references.}
\label{fig:exp5}
\end{figure*}

% p1=p2=p3=0.33
\subsection{Optimal design for individual preference}

We demonstrated that CoCoG-2 can directly generate images guided by behavioral results, significantly facilitating the rapid production of desirable experimental outcomes. Beyond this, CoCoG-2 can also be used to design and generate visual stimuli that maximize information gain, thereby substantially reducing the number of experimental trails required in cognitive research. Taking individual preferences in the {\it Two alternative forced choice} experiment as an example, for the same set of inference images and query image, each participant may choose a different inference image as the more similar one. The greater the variance in participants' choices, and the more information each experiment can provide about individual preferences. Therefore, when the entropy of the probability distribution for choosing between the two reference images is maximized, each {\it Two alternative forced choice} experiment provides the maximum information gain.

Furthermore, we consider each experiment's three images from the perspective of the {\it Odd One Out} experiment. The Odd One Out experiment is a triplet choice task where participants select the image that is least similar to the others as the odd one out. This experiment can reveal the similarity relationships between the three images, and its results can be represented by the probability of each image being selected as the odd one out. Therefore, when the entropy of the probability distribution for the three images being chosen as the odd one out is maximized, each Odd One Out experiment provides the maximum information gain.

Consequently, our optimization goal is to maximize the entropy of the probability distribution for the three images being chosen as the odd one out, guided by {\it Uncertainty guidance}. The choice probability is determined by the similarity between the images, which in turn is dictated by their relative distances in Concept Space. Maximizing the entropy of the probability distribution implies that the three images form an equilateral triangle in Concept Space, as shown in Figure~\ref{fig:exp5}a. Since Concept Space has 42 dimensions, there can be at most 43 possible query images that meet the equilateral triangle condition. Figure~\ref{fig:exp5}c displays generated images corresponding to three pairs of reference images. These generated images are positioned at the vertices of multiple equilateral triangles in the 42-dimensional space, exhibiting good diversity. Additionally, Figure~\ref{fig:exp5}b shows the probability distribution for these three sets of images, with the distribution centers very close to the center of the equilateral triangle, indicating that the generated images well satisfy the target probability distribution. These images are diverse and informative, making them optimal experimental images for studying individual preferences.

% given x & y, find c (post explain); given x, change c (reason), expect y (result) 
% \subsection{concept counterfactual reasoning}

% \subsection{Related works}

% \subsection{Promoting the Alignment of AI and Human in High-level Visual Cognition}

% \subsection{Facilitating Optimal Design}

\section{Discussion}

In conclusion, our proposed CoCoG-2 framework advances the field of controllable visual object generation by integrating concept representations with behavioral outcomes to guide the image generation process. By employing a versatile experiment designer and utilizing meticulously designed guidance, CoCoG-2 not only addresses the limitations of its predecessor, CoCoG, but also significantly enhances the flexibility and efficiency of generating visual stimuli for cognitive research. The experimental trials generated using CoCoG-2 are instrumental in probing the causal relationships between concept representations and human decision-making behaviors. This approach marks a substantial step towards understanding and manipulating concept-based representations in a controllable and meaningful way, offering a robust methodology for cognitive science research and practical applications in AI-driven visual content generation.

\section*{Ethical Statement}
The human behavioral data in this study is from THINGS public dataset. No animal or human experiments are involved.

\bibliographystyle{splncs04}
\bibliography{named}

\end{document}